\documentclass[conference,a4paper]{IEEEtran}
\usepackage[T1]{fontenc}
\usepackage{amsmath,amssymb,booktabs,array,graphicx,cite}
\usepackage{microtype}
\usepackage{balance}
\usepackage[hidelinks]{hyperref}
\hypersetup{pdftitle={Memory Admission and Result-Retention Sizing in an RV32IMC FPGA Soft Processor},pdfauthor={Yihang Dai, Shuyu Xu, Yinfeng Zhou, Chenglei Peng}}
\usepackage{url}
\usepackage{eso-pic}
\graphicspath{{figures/}}
\newcommand{\cmhz}{CoreMark/MHz}
\newcommand{\eff}{CoreMark/MHz/kLUT}
\begin{document}
\title{Memory Admission and Result-Retention Sizing in an RV32IMC FPGA Soft Processor}
\author{\IEEEauthorblockN{Yihang Dai}
\IEEEauthorblockA{Nanjing University\\Nanjing, China\\241180155@smail.nju.edu.cn}
\and
\IEEEauthorblockN{Shuyu Xu}
\IEEEauthorblockA{Nanjing University\\Nanjing, China\\241180151@smail.nju.edu.cn}
\and
\IEEEauthorblockN{Yinfeng Zhou}
\IEEEauthorblockA{Nanjing University\\Nanjing, China\\241180112@smail.nju.edu.cn}
\and
\IEEEauthorblockN{Chenglei Peng\IEEEauthorrefmark{1}}
\IEEEauthorblockA{Nanjing University\\Nanjing, China\\pcl@nju.edu.cn\\\IEEEauthorrefmark{1}Corresponding author}}
\maketitle
\AddToShipoutPictureFG*{\AtPageLowerLeft{
  \put(0,14){\makebox[\paperwidth][c]{%
    \parbox[b]{\textwidth}{\fontsize{7}{8}\selectfont \textcopyright\ 2026 IEEE. Personal use of this material is permitted. Permission from IEEE must be obtained for all other uses, in any current or future media, including reprinting/republishing this material for advertising or promotional purposes, creating new collective works, for resale or redistribution to servers or lists, or reuse of any copyrighted component of this work in other works.}%
  }}
}}
\raggedbottom

\begin{abstract}
In FPGA soft processors with shared backing block RAM, memory admission can stall a multiplication after its product has been prepared. We study how admission timing changes result retention in PaceRV, our board-tested five-stage RV32IMC core, using fixed binaries and matched dynamic instructions. Offering data requests one edge earlier reduces geometric-mean Embench cycles by 18.2\% and retention opportunities from 375,063 to 4,373. Admission and retention cycle gains are conditional and non-additive. Under slower reads, a 99\% capture target selects an 18-edge validity window with registered requests and 28 edges with early requests, although no multiplication matched within both measurement intervals waits longer. A predeclared sizing rule predicts opportunity recovery exactly in all finite tests, including kernels excluded from model development. Independent timing analysis closely estimates execution-cycle savings. The results support choosing admission first, then deciding whether to extend validity and sizing it against an explicit capture or missed-opportunity objective.
\end{abstract}


\section{Introduction}
Synchronous block RAM shapes request and response timing in FPGA processors. Frontend and load/store accesses share backing memory, and interface registers can postpone request acceptance. In a short in-order pipeline, this extra edge can also delay a younger multiplication whose product is already prepared.

PaceRV is our single-issue, in-order RV32IMC core. An eligible multiplication prepares its low-word product on entering the execute (EX) stage, but an older load stalled in the memory stage can delay its use. Each prepared product belongs to the dynamic MUL instruction occupying EX; the measured population comprises successive multiplications in one core. A one-edge valid flag allows the fast path to accept the product at the next edge. If the stall outlasts this window, ordinary multiplication remains available. Retention extends the window, while earlier memory admission can remove the wait.

The design question is whether to extend result validity and, if so, for how long. We track the same dynamic instructions across admission modes to answer it, separating recovered opportunities from elapsed-cycle savings.

Shared multipliers require retention policies for prepared work held by a stalled requester. Admission-conditioned demand and validity windows provide inputs for selecting retention and result-storage policies between cores.

Earlier admission moves most delayed multiplications into immediate use. Absolute demand falls sharply, but longer waits form a larger share of the remainder. A fixed percentage target can consequently select a longer validity window despite a smaller application-level benefit. This changes the design choice: establish admission timing before extending validity, and choose the window using the remaining demand and the intended performance objective.

PaceRV includes caches, shared backing RAM, machine services and live counters. Fixed binaries and retirement matching isolate admission and validity effects. The instruction-resident suite chiefly exposes request-register latency; direct instruction/data command conflicts are rare.

The study makes three contributions. First, instruction-level age measurements explain a population shift: early admission removes 370,690 short waits, causing fractional-capture and absolute-demand criteria to select different validity windows. Second, a predeclared sizing procedure turns the remaining demand into a validity-window choice and predicts opportunity recovery exactly across 152 finite executions, changed read schedules and four new kernels. Independent timing measurements quantify endpoint and cycle accuracy. Third, our five-stage RV32IMC implementation connects the analysis to routed resource cost and board performance. PaceRV retains full peripherals and 64-bit counters, reaches 3.442 \cmhz\ at 145.125 MHz, and supports comparison with a feature-matched VexRiscv system.

\section{Background and Problem Formulation}
\label{sec:background}
The \emph{admission edge} is when the memory interface accepts a request; its response arrives later. Registering the address adds an edge even with unchanged cache and backing service. FPGA memory studies distinguish organization and data movement from application cycles~\cite{kamaleldin2019modular} and examine cache/circuit tradeoffs in out-of-order processors~\cite{wong2016memory}. Processor customization trades frequency against cycles~\cite{yiannacouras2007exploration}, including five-stage RISC-V designs tuned for memory and compressed instructions~\cite{miyazaki2020rvcorep,kanamori2021rvcorep32ic}. Per-core request registers and round-robin multiplier/divider service extend these resource constraints to shared multicores~\cite{islam2022sharing}. We examine how admission timing changes retention demand and the required validity window.

A \emph{prepared multiplication} is a dynamic low-word multiply (MUL) instruction whose product and valid flag are stored with its EX-stage state. Consumption means acceptance of this product by the fast path. Its age is the preparation-to-use interval in rising clock edges, $a_i=c_i-p_i$, where $p_i$ is preparation on EX entry and $c_i$ is consumption. The \emph{validity window} $K$ permits fast use when $1\leq a_i\leq K$. Thus $K=1$ permits use at the next edge and $K=\infty$ keeps the product valid until use or cancellation. This window governs fast-path use, independently of the memory admission edge. Expiry discards neither the instruction nor its memory request; ordinary multiplication still produces the architectural result. A flush or exception cancels the instruction, and EX departure clears its valid flag.

Prepared-result retention preserves work for the same dynamic instruction while it remains in EX. Latency-insensitive and synchronous elastic design preserve valid data under backpressure~\cite{carloni2001theory,cortadella2006elastic}, and tags separate producer identity from values~\cite{tomasulo1967algorithm}. Dynamic instruction reuse stores results for later instruction instances~\cite{sodani1997reuse}. Here the product remains associated with the same instruction.

Interaction-cost analysis~\cite{fields2003interaction}, cycles-per-instruction (CPI) accounting~\cite{eyerman2006counter,eyerman2018multistage} and mechanistic models~\cite{karkhanis2004firstorder,breughe2012inorder} explain why overlapping events do not add up independently in elapsed cycles. Our replay follows each multiplication's preparation and consumption, checking absolute edges and their difference. Phased scratchpad deployment for multicore real-time systems~\cite{dudzik2025phased} also models memory-service constraints, but schedules tasks rather than instruction-level result validity.

\section{Processor and System Architecture}
\label{sec:architecture}
\begin{figure}[t]
\centering\includegraphics[width=\columnwidth]{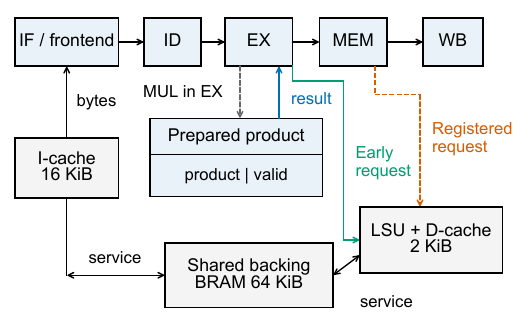}
\caption{Admission and preparation in PaceRV. Registered requests add an address-register edge; early requests are offered one edge sooner. The prepared product and valid flag belong to the instruction in EX. Both caches share backing BRAM.}
\label{fig:architecture}
\end{figure}

PaceRV's five stages are instruction fetch (IF), decode (ID), execute (EX), memory access (MEM) and writeback (WB). Its RV32IMC instruction set supports integer, multiply/divide and compressed instructions. Instructions issue and retire in order, one at a time. The frontend and load/store unit (LSU) have separate caches backed by shared block RAM (BRAM). With \emph{registered requests}, a data request is stored before being offered one edge later. With \emph{early requests}, it is offered on EX-to-MEM transfer (Fig.~\ref{fig:architecture}). Older-request priority, exception suppression and acceptance tracking prevent duplicate or out-of-order admission.

\begin{figure}[t]
\centering\includegraphics[width=\columnwidth]{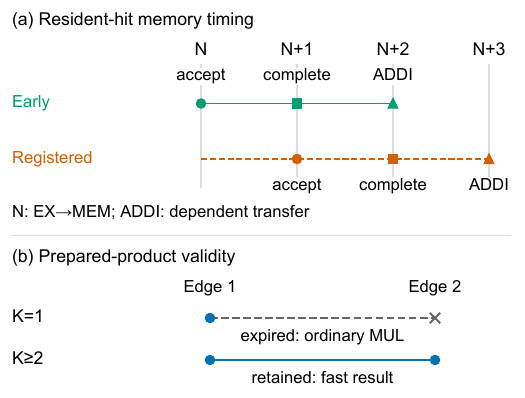}
\caption{Memory admission and fast-path validity. (a) Directed resident hit, with $N$ the EX-to-MEM transfer edge. (b) Product preparation is age zero; validity window $K$ counts subsequent edges on which fast consumption is permitted. Expiry leaves ordinary MUL completion available.}
\label{fig:timing}
\end{figure}

In the uncontended resident-hit test, the early mode accepts the request at edge $N$ and completes it at $N+1$; the registered mode accepts it at $N+1$ and completes it at $N+2$. A dependent add-immediate (ADDI) instruction advances one edge after completion in both settings. Figure~\ref{fig:timing} separates this schedule from prepared-product validity.

Prepared arithmetic stays bound to the MUL in EX. Its low-word product is prepared on entry when operands are available, no unresolved older writer remains and the multiplier is ready. When memory backpressure clears, a valid prepared product lets the MUL leave EX directly. The ordinary path first captures partial products and supplies the result one edge later, so retention preserves a local one-edge advantage through the stall. The baseline valid flag lasts one edge; retention protects the stored product until fast consumption is accepted. EX departure clears the old flag. An exception, an older MEM flush, a frontend barrier or reset cancels the product, while an ID-only redirect can preserve older EX work. Cancellation takes priority over admission; clearing a departing instruction's flag preserves a newly arriving instruction's flag.

A finite $K\geq2$ needs $\lceil\log_2 K\rceil$ countdown bits plus a hold bit that protects the stored product after validity expires, until multiplier-request acceptance, EX release or cancellation. Expiry clears the valid flag; it does not release product storage. The selected 18- and 28-edge windows both use five countdown bits. Unbounded validity needs neither the countdown nor its additional hold bit. All modes keep the same preparation conditions and memory capacity.

\section{Predicting Retention Demand Under Changed Service}
\label{sec:model}
\subsection{Retention demand and validity window selection}
For a common set of multiplications $\mathcal O$ consumed on the fast path with unbounded validity in the measurement interval, let $a_m(i)$ be instruction $i$'s age at use under request mode $m$. Additional demand excludes age-one use already served by the baseline:
\begin{align}
N_m &= \sum_{i\in\mathcal O}\mathbf1[a_m(i)>1],\label{eq:demand}\\
R_m(K) &= \sum_{i\in\mathcal O}\mathbf1[a_m(i)>K],\label{eq:remaining}\\
F_m(K) &= 1-R_m(K)/N_m,\qquad N_m>0.\label{eq:capture}
\end{align}
Here $N_m$ counts multiplications whose prepared products must remain valid beyond the next edge. Of these, $R_m(K)$ require more than $K$ edges and lose the fast-path opportunity; $F_m(K)$ is the fraction served within the window. The target $q$ is the desired fraction of these delayed multiplications served, set to 0.99 in this study. We select the smallest integer $K\geq2$ with $F_m(K)\geq q$. When $N_m=0$, no additional retention is needed and capture is undefined.

The two measures behave differently when admission changes. If every multiplication is consumed no later under early requests, $a_{\rm early}(i)\leq a_{\rm reg}(i)$, each indicator in~\eqref{eq:remaining} can only shrink, so $R_{\rm early}(K)\leq R_{\rm reg}(K)$ for all $K$ and the validity window needed for a fixed budget of missed opportunities cannot grow. $F_m(K)$, however, divides by a smaller $N_m$. Removing many short waits leaves the long ones as a larger share, and the selected quantile moves out. Matching multiplications across settings establishes the connection between this population shift and the same instructions' changed consumption ages.

The prospective protocol fixed $q=0.99$, the selection rule and the validation points before slower-service profiling. Numerical predictions from the unbounded profiles were then fixed before finite execution. Each mode is tested at one edge, the selected validity window, the integer below it and the maximum observed age.

Cycle savings need a model of their own. Let $w_m(i)$ equal one when ID admits another instruction at multiplication $i$'s consumption edge in the unbounded profile for mode $m$, and zero otherwise. The local cycle estimate is
\begin{equation}
\widehat S_m(K)=\sum_{i\in\mathcal O}\mathbf1[1<a_m(i)\leq K]w_m(i),
\label{eq:cycle-estimate}
\end{equation}
which assumes that later frontend and memory phases preserve each local saving. Measured savings decompose as $S=O-\Delta_{\rm occ}$: one cycle per additional consumption $O$ minus an occupancy residual $\Delta_{\rm occ}$. We measure this residual $\Delta_{\rm occ}$ after execution.

\subsection{Timing prediction under changed memory service}
To forecast a changed read schedule, we build a graph from the original execution. Nodes represent accepted pipeline transfers, requests, responses and backing-memory beats. Edges constrain when a pipeline stage becomes available and when a shared resource can serve the next event. The recorded preparation flag identifies multiplications with prepared products. The service change adds delay only to the affected read-beat edges, and replay keeps the recorded event order and cache outcomes.

The replay distinguishes memory delay from peripheral and machine-control timing. Frontend delivery follows a redirect by at least one edge. A memory-mapped input/output (MMIO) access takes two edges to respond through PaceRV's registered peripheral interface. A control/status register (CSR) access or machine return (MRET) leaves EX at least one edge after the last older retirement, assuming no interrupt is being serviced. Exception cancellation follows Section~\ref{sec:architecture}. The universal asynchronous receiver/transmitter (UART) is the serial peripheral, whose readiness follows an absolute deadline.

Stage and resource constraints determine a lower bound on each event's edge. Waiting beyond this bound is represented by a source-trace residual and propagated under the changed schedule. Let $P(v)$ contain event $v$'s prerequisites, including these interface constraints, and let $\ell_{uv}$ be their minimum delays under the original service. For source event edges $t_v$, define the causal minimum $b_v$ and residual $g_v$:
\begin{equation}
b_v=\max\bigl(\{0\}\cup\{t_u+\ell_{uv}:u\in P(v)\}\bigr),\quad g_v=t_v-b_v.
\label{eq:gap}
\end{equation}
The source residual $g_v\geq0$ is computed before the service change. Declared read-service increments $\delta_{uv}$ then give the target causal minimum
\begin{equation}
B_v=\max\bigl(\{0\}\cup\{\widehat t_u+\ell_{uv}+\delta_{uv}:u\in P(v)\}\bigr).
\label{eq:replay}
\end{equation}
Dependency-relative propagation carries each event's original residual through the changed prerequisites. For nonroot events other than UART data-register responses, it uses $\widehat t_v=B_v+g_v$. Roots and UART data-register responses retain an absolute timing anchor: they use $\max(B_v,t_v)$ when $g_v>0$, and $B_v$ otherwise. Replay with unchanged service reproduces the source trace. The model applies where relevant instruction ordering and survival are preserved; we assess these conditions through command order, cache outcomes, instruction survival and architectural values, and measure prediction accuracy at preparation and consumption.

The predicted preparation and consumption edges give each multiplication's age. Moving both endpoints by the same amount changes absolute timing but not the preparation-to-use interval, so endpoint and interval accuracy are evaluated separately. Unchanged endpoints provide a null comparator. Absolute propagation instead releases unexplained events at their original edges. A targeted service test crosses recorded or interface-refined minimum delays with absolute or dependency-relative propagation. All four variants use the same source events, target service and prepared-instruction set.

\section{Evaluation}
\label{sec:evaluation}
\subsection{Experimental setup and comparability}
The target is a Pango Logos2 PG2L200H-6FBB676 FPGA with Pango Design Suite 2022.2-SP6.4 and ADS synthesis. Logic cost uses lookup tables (LUTs) and flip-flops (FFs); native Pango RAM and arithmetic blocks are denoted DRM and APM. Board deployments use early admission: PaceRV Base has $K=1$ and PaceRV Retained has $K=\infty$. Table~\ref{tab:configuration} defines the system boundary. Nineteen pinned Embench programs~\cite{embench_source} use identical RV32IMC binaries, warmup and global scale one; a measurement dummy is excluded. All programs enter the aggregates. CoreMark~\cite{coremark_source} uses GCC 10.2.0, RV32IMC/ILP32, \texttt{-O3}, loop unrolling, 2,000 data bytes and 6,000 iterations lasting over ten seconds. Firmware uses fixed freestanding/linker settings and clock-matched UART configuration. The processor implementation, experimental data and reproduction scripts are publicly available.\footnote{\url{https://github.com/BXSYGJ/PaceRV-MCSoC2026}}

\begin{table}[t]
\caption{System configurations and measurement scopes. VexRiscv uses revision \texttt{680756065e9e} (11 February 2026).}
\label{tab:configuration}
\centering\footnotesize
\setlength{\tabcolsep}{3pt}
\begin{tabular}{@{}>{\raggedright\arraybackslash}p{0.23\columnwidth}>{\raggedright\arraybackslash}p{0.70\columnwidth}@{}}
\toprule
Property & Configuration or evidence scope\\\midrule
Common memory & 16 KiB I-cache, 2 KiB D-cache; two ways, 16-byte lines; shared 64 KiB backing RAM.\\
Common services & RV32IMC; cycle/retired CSRs; writable \texttt{mtvec}/\texttt{mscratch}; ECALL, interrupts, UART, GPIO, timer, atomic 64-bit snapshots.\\
Monitoring & Live native cache events in both; Vex dirty-line writebacks are zero under its write-through policy.\\
Cache policies & PaceRV write-back versus Vex write-through/nonallocating stores; replacement, cache pipeline, delivery and event semantics remain native.\\
Vex execution & EX branch resolution; 256-entry dynamic-target predictor; native bypassing and multiply/divide.\\
Validation & PaceRV deployments: three cold board runs per CoreMark role. Vex: two cold RTL runs per role, routed timing/resources.\\
\bottomrule
\end{tabular}
\end{table}

For iterations $I$, measured cycles $C$ and full-system packed LUT count $L$, we report $\cmhz=10^6I/C$ and $\eta_{\rm LUT}=10^9I/(CL)$ in \eff. Packed LUT occupancy includes distributed RAM once; BRAM and arithmetic blocks are separate. Throughput uses the configured clock. Measured intervals include snapshot overhead but not UART reporting.

Embench geometric means weight programs equally. For control/default cycle ratios $r_w$, $G=\exp(\sum_w\log r_w/19)$; we report cycle reduction $1-1/G$, which is smaller than the same-clock throughput gain $G-1$.

For an external implementation reference, we build a VexRiscv~\cite{vexriscv_source} system with the same cache geometry, machine services and backing infrastructure as PaceRV, retaining its native policies. Comparison studies motivate explicit system boundaries~\cite{goens2016mapping,doerflinger2021survey}; these are specified in Table~\ref{tab:configuration}. Vex produces valid CRCs, and its Embench marker streams match PaceRV's.

PaceRV Base and Retained pass fifty instruction-set and system tests and the public suite, covering expiry, hazards, backpressure, exceptions, interrupts, redirects and reuse. The compared measurement intervals have identical ordered PC/instruction digests and retirement counts. Joined multiplications agree in retirement ordinal, operands, destination and product. Passive taint tracking finds no counter-derived address or branch in a measured interval. Admission, validity window and read-service interventions use unchanged binaries in register-transfer-level (RTL) simulation; physical implementation and board measurements establish their operating context separately.

\subsection{Conditional effects of memory admission}
When the request register is moved so that data requests leave one edge earlier, geometric-mean Embench cycles drop by 18.2\%, almost entirely through shorter admission latency. In the matrix profiles, early requests are accepted on EX-to-MEM transfer and registered requests one edge later, with the same acceptance-to-response latency. Lookup, refill and writeback traffic is unchanged, and the pooled D-cache hit rate is 95.0\%. Port conflicts are rare: across the suite, data commands wait behind instruction fetch for fewer than 600 cycles in either mode.

Early admission increases immediate consumption of prepared products in two programs. Both settings prepare 312,000 multiplications in \texttt{matmult-int}, and early requests raise immediate consumption from 124,800 to 308,022. In \texttt{edn}, immediate consumption rises from 416,105 to all 518,400 preparations. Because \texttt{Xgboost} prepares no multiplication, its 891,078-cycle reduction isolates the admission effect.

The retention experiment keeps preparation, prediction and fetch settings fixed, varying admission and validity window. Conditional admission and retention gains are non-additive.

\begin{figure*}[t]
\centering\includegraphics[width=\textwidth]{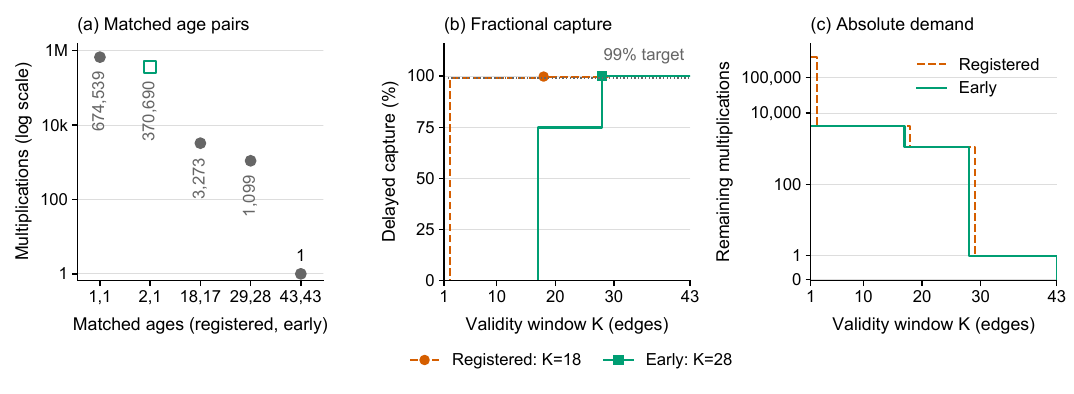}
\caption{Earlier admission moves 370,690 multiplications from delayed to immediate use under slower reads. (a) Age pairs of 1,049,602 multiplications used inside both measurement intervals. (b) A 99\% target selects $K=18$ with registered requests and $K=28$ with early requests from delayed populations of 375,063 and 4,373. (c) Remaining demand never increases with early requests. Panel (a) is logarithmic; panel (c) is linear from zero to one and logarithmic above one.}
\label{fig:population}
\end{figure*}

\begin{table*}[t]
\caption{Admission, sizing and forecasting results. An opportunity is fast product use beyond age one. MAE denotes mean absolute error.}
\label{tab:results}
\centering\footnotesize\setlength{\tabcolsep}{5pt}
\begin{tabular}{@{}llrrl@{}}
\toprule
Service / test & Quantity & Registered & Early & Interpretation\\\midrule
Original & Additional opportunities & 375,063 & 4,373 & Seven / two of nineteen programs benefit\\
Original & Cycles saved by retention & 374,973 & 4,373 & Full-suite reductions: 0.539\% / 0.0091\%\\\midrule
Slower reads & Selected 99\% validity window $K$ & 18 & 28 & Two added edges before every read beat\\
Slower reads & Capture at selected $K$ (\%) & 99.7067 & 99.9771 & Below-target neighbors: $K=17$ / $K=27$\\
Finite execution & Exact opportunity predictions & 76/76 & 76/76 & Nineteen programs, four windows per mode\\
Finite execution & Cycle prediction error (cycles) & 80 & 0 & Each nonbaseline window; forecast minus measurement\\
Leave-one-out & Held-out programs meeting 99\% & 5/7 & 2/2 & Pooled held-out capture 98.9\% / 99.98\%\\\midrule
New bit-6/7 service & Endpoint mean absolute error (edges) & 0.595 / 0.595 & 0.595 / 0.595 & Preparation / consumption; frozen predictions\\
New bit-6/7 service & Exact paired preparation-to-use intervals & 1,088,536 & 1,088,536 & Same nineteen programs; unbounded target\\\midrule
New bit-8/9 service & Selected 99\% validity window $K$ & 12 & --- & \texttt{matmult-int}; measured capture 99.0833\%\\
New bit-8/9 service & Multiplications with both endpoints exact & 320,000 & --- & Unbounded target; relative propagation\\\midrule
Four new kernels & Matmul opportunities at $K=2$ & 1,538 & 0 & Per schedule; all recovered; 1,534 cycles saved\\
Four new kernels & Matmul endpoint MAE (edges) & 0 & 0 & Absolute release 34.0--36.0; intervals exact\\
\bottomrule
\end{tabular}
\end{table*}

\subsection{Residual retention demand}
Under original memory service, unbounded retention recovers 375,063 additional fast consumptions with registered requests but only 4,373 with early requests (Table~\ref{tab:results}). Seven programs benefit in the registered mode. With early requests only \texttt{matmult-int} and \texttt{picojpeg} benefit, and the suite-wide cycle reduction falls from 0.539\% to 0.0091\%. Retention therefore provides little additional runtime benefit after early admission. Registered-mode consumptions save 374,973 cycles, 90 fewer than their count; occupancy accounting attributes the difference to empty frontend and EX slots. With early requests each consumption saves one cycle.

Slower reads add two wait edges before every beat, increasing the ideal interval from one to three edges. Writes, memory geometry and binaries remain unchanged. Both services have identical measured delayed-multiplication sets in each mode (375,063 with registered requests and 4,373 with early requests), while slower reads change their longer-wait ages. Matching retired instructions across the complete execution identifies 1,088,536 prepared multiplications, all consumed on the fast path in both modes. Figure~\ref{fig:population} uses the 1,049,602 consumed inside both measurement intervals; the remaining 38,934 are fast consumers outside both. None of the plotted multiplications waits longer under early requests: 674,539 stay at age one, 370,690 move from age two to age one, and the rest shift down by one edge or stay at 43.

The age-two group accounts for 98.8\% of registered-mode additional demand. At $K=17$ missed opportunities drop from 4,373 with registered requests to 1,100 with early requests, while capture drops from 98.8\% to 74.8\%: the denominator shrinks faster than the tail. A 28-edge early-mode window captures 4,372 of 4,373 opportunities and meets the 99\% target. This window represents a residual-demand quantile, rather than a runtime optimum. An absolute budget of 1,100 misses instead selects 18 edges with registered requests and 17 with early requests. Window selection therefore depends on whether the objective is fractional capture or absolute demand.

\begin{figure*}[t]
\centering\includegraphics[width=\textwidth]{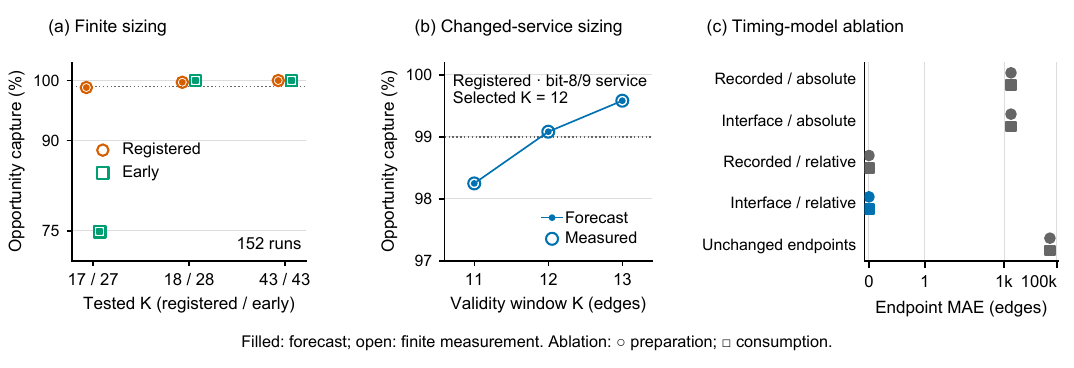}
\caption{Sizing and timing prediction. (a) Forecasts match capture in 152 slower-read finite executions. (b) The frozen bit-8/9 forecast selects $K=12$ for \texttt{matmult-int} with registered requests, confirmed by adjacent windows. (c) The 320,000-multiplication ablation crosses recorded/interface-refined minimum delays with absolute/relative propagation. Both relative variants have zero endpoint MAE. The error axis is linear near zero and logarithmic above 0.1 edge.}
\label{fig:prediction}
\end{figure*}

\subsection{Finite-window validation and workload transfer}
Finite execution tests whether an unbounded multiplication profile predicts recovery within a finite window. The predeclared rule chooses $K=18$ for registered requests and $K=28$ for early requests. Nineteen programs at four windows per mode give 152 executions: one edge, the selected window, the integer below it and the maximum age of 43. Recovery predictions match every execution; capture first reaches 99\% at the selected windows (Table~\ref{tab:results}). Across 8,708,288 finite--unbounded comparisons, instruction identities, operands, writebacks and fast-consumer membership agree.

Cycle forecasts closely track finite execution. With registered requests at $K=18$, 373,963 recovered opportunities yield 373,953 predicted and 373,873 measured saved cycles. The local model accounts for ten occupancy cycles; the remaining eighty-cycle error reflects the approximation that later phases preserve each saving. This is 0.0214\% of predicted savings in a 94,800,126-cycle execution. Across the three nonbaseline registered-mode windows the error is at most 0.022\%; early-mode estimates match (Fig.~\ref{fig:prediction}).

We also asked how well a validity window chosen on some programs carries over to others. Holding out one program at a time, we rerun the selection on the frozen unbounded profiles of the other eighteen. With early requests every fold selects $K=28$, and both programs with residual demand reach 99\%. With registered requests five of seven reach it, and pooled held-out capture is 98.9\%. One of the two misses is the program with the most demand. When \texttt{matmult-int} is held out, the remaining demand is almost all age two, the rule picks $K=2$, and \texttt{matmult-int} itself reaches 97.9\%. \texttt{Picojpeg} has a 29-edge tail of its own and reaches 85.0\% at $K=18$. The rule carries over as long as the profile contains tail ages like those of the new program.

\subsection{Forecasting changed memory service}
We test whether a graph built from the original execution predicts preparation and consumption times under a changed read schedule. The first target schedule latches the accepted command address. Its first accepted beat follows command acceptance by $2+2\cdot\mathbf1[\mathrm{addr}[6]]$ edges; later accepted beats are spaced by $1+\mathbf1[\mathrm{addr}[7]]$ edges. Both conditions may apply, and writes are unaffected. Endpoint forecasts were frozen before target compilation or profiling. The same nineteen programs in both request modes provide 38 unbounded targets.

Propagating each event's source residual relative to its prerequisites gives a mean absolute error (MAE) of 0.595 edges for preparation and consumption separately over 1,088,536 multiplications per mode. Each endpoint forecast is compared directly with its target trace; instruction matching establishes identity without adjusting timestamps. Absolute propagation has an endpoint MAE of about 6,000 edges, and unchanged endpoints about 28,000. All 2,177,072 paired preparation-to-use intervals match: residual errors of at most eight edges, confined to three programs, shift both endpoints together. The forecast also tracks the 99\% window; \texttt{matmult-int} with registered requests moves from 10 to 12 edges.

Memory command sequences match in 37 of 38 cases. With early requests in \texttt{sglib-combined}, one four-beat data read overtakes ten others, including another data read, outside the measured interval; the command multiset and instruction-master order are preserved. This program prepares no multiplications and contributes no preparation-to-use interval comparisons.

A second schedule separates the contribution of residual propagation from that of interface minimum delays. For \texttt{matmult-int} with registered requests, the first beat is delayed by $2+2\cdot\mathbf1[\mathrm{addr}[8]]$ edges and later beats by $1+\mathbf1[\mathrm{addr}[9]\oplus\mathrm{addr}[8]]$. Forecasts and windows $K=1,11,12,13$ were fixed before the unbounded and four finite target executions. The forecast selects $K=12$ for 187,200 delayed multiplications; measured capture is 98.25\%, 99.0833\% and 99.5833\% at the three nonbaseline windows. All four graph variants select the same validity window and predict the same capture. At $K=12$ retention saves 185,481 cycles, with a two-cycle overestimate at each nonbaseline window.

In the unbounded target, both relative-propagation variants predict preparation and consumption exactly for all 320,000 multiplications. Absolute propagation has 1,836.95-edge endpoint MAE and unchanged endpoints about 54,621.5 edges (Fig.~\ref{fig:prediction}). Relative propagation accounts for the endpoint improvement; refining interface delays provides no additional improvement on this workload. Validation covers the prepared-instruction prefix, where retirement order, backing commands and data-cache outcomes match. Post-measurement UART reporting and $K=1$ frontend request multiplicity vary.

\begin{table*}[t]
\caption{Complete systems at 145.125 MHz. PaceRV CoreMark uses board cycles; Vex uses RTL cycles. BRAM/DSP counts mean native Pango DRM/APM blocks. Setup slack includes 0.150 ns uncertainty.}
\label{tab:physical}
\centering\footnotesize\setlength{\tabcolsep}{4pt}
\begin{tabular}{@{}lrrrrrrrrrr@{}}
\toprule
Implementation & MHz & LUT & FF & BRAM & DSP & Setup (ns) & CoreMark & CM/MHz & $\eta_{\rm LUT}$ & \shortstack{Cycle\\source}\\\midrule
PaceRV Base & 145.125 & 6,097 & 2,541 & 25 & 4 & +0.047 & 499.506 & 3.441903 & 0.564524 & Board\\
PaceRV Retained & 145.125 & 6,150 & 2,541 & 25 & 4 & +0.016 & 499.531 & 3.442075 & 0.559687 & Board\\
Vex EX, matched & 145.125 & 4,152 & 2,313 & 25 & 4 & +0.474 & 398.007 & 2.742514 & 0.660528 & RTL\\
\bottomrule
\end{tabular}
\end{table*}

\subsection{Transfer to new workloads}
Workload transfer tests the frozen model on four public RISC-V benchmarks~\cite{riscvTestsIbexPinned} excluded from its development. The ports preserve algorithms, inputs and output checks: $16\times16$ integer matrix multiplication, quicksort of 2,048 integers, median of 400 integers and addition of 300-element vectors. Each firmware supplies an original-service profile, and target predictions are fixed before execution on the bit-6/7 and bit-8/9 schedules. The 58 executions comprise eight source profiles, sixteen unbounded targets and thirty-four finite runs.

Across both schedules, the forecast selects $K=2$ for matrix multiplication's 1,538 delayed multiplications under registered requests, and finite execution recovers all of them. The model predicts 1,538 saved cycles against 1,534 measured, a four-cycle overestimate (0.26\% of the estimate, 0.0086\% of elapsed cycles); early requests eliminate residual demand. Both endpoints of all 16,396 paired multiplications are predicted exactly, compared with 34--36-edge MAE for absolute release and about 1,265 for unchanged endpoints. Direct comparison of timestamp streams confirms endpoint accuracy before computing preparation-to-use intervals. All models predict these intervals correctly; absolute endpoints distinguish them. Quicksort, median and vector addition supply zero-demand controls with the predicted zero recovery.

\subsection{Physical implementation and external comparison}
\begin{figure}[t]
\centering\includegraphics[width=\columnwidth]{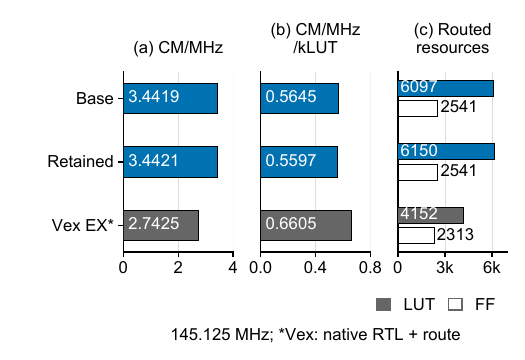}
\caption{Execution efficiency, LUT efficiency and routed occupancy; CM denotes CoreMark. PaceRV uses board-qualified cycles; Vex uses RTL cycles. Table~\ref{tab:physical} reports memory, arithmetic blocks and timing.}
\label{fig:implementation}
\end{figure}

Table~\ref{tab:physical} and Fig.~\ref{fig:implementation} separate execution efficiency, resources and timing at the common clock.
PaceRV Retained uses 53 more LUTs than Base (6,150 versus 6,097), with identical FF, BRAM and arithmetic-block counts. This difference includes whole-netlist mapping and placement. Both PaceRV implementations in Table~\ref{tab:physical} satisfy all ten timing checks at 145.125 MHz. Setup slack includes 0.150 ns clock uncertainty; CoreMark cycle measurements come from the board for PaceRV and RTL simulation for Vex.

PaceRV reaches 3.442 \cmhz\ at 145.125 MHz. Three cold board runs per CoreMark configuration reproduce the corresponding RTL results, with matching CRCs, cycle counts, retirements and UART output. Base and Retained have identical retirement counts and nearly identical execution cycles, consistent with the small residual demand under early admission.

The routed finite implementations also meet timing. Under original service, matched registered-mode routes use 6,003 LUTs for unbounded retention, 6,033 at $K=2$ and 6,058 at $K=10$. An early-mode route at $K=20$ uses 6,164 LUTs and captures 4,372 of 4,373 opportunities. Routed occupancy reflects countdown and hold logic together with whole-netlist mapping.

The matched systems separate execution efficiency from LUT efficiency (Fig.~\ref{fig:implementation}). PaceRV achieves 25.5\% higher \cmhz, while VexRiscv uses 4,152 LUTs and reaches 0.661 \eff. Over the nineteen Embench images Vex needs 6.1\% more cycles in geometric mean (4.6\% without \texttt{nsichneu}), and PaceRV is faster on thirteen programs. These results describe the complete systems as tested, including their native write and delivery policies.

\section{Discussion}
\label{sec:discussion}
The findings change whether validity should be extended and how its window is chosen. Admission comes first: early requests leave seventeen of nineteen measured programs with no additional retention demand. A fractional target sizes coverage of the remainder; an absolute budget bounds missed opportunities. Under slower reads, 99\% capture selects 18 edges with registered requests and 28 with early requests. Allowing 1,100 misses instead selects 18 and 17 edges. The longer percentage-selected window therefore expresses a different objective, even though matched multiplications consume their results no later.

Instruction-level analysis predicts cycle savings by identifying usable prepared results and estimating their contribution. The observed eighty-, two- and four-cycle overestimates quantify the error introduced by assuming that later phases preserve each local saving. Each recovered fast consumption supplies a potential cycle saving; frontend availability and later overlap determine the measured reduction.

Preparation and consumption are distinct timing targets because correlated endpoint errors cancel in their difference. Dependency-relative propagation provides accurate absolute forecasts across the tested schedules. Transfer requires a source profile and the processor's interface minimum delays, with relevant instruction order and survival preserved. The new-kernel study includes one workload with retention demand and three zero-demand controls. Profile-free prediction, changed control flow and richer cancellation behavior are directions for future work.

Shared multipliers~\cite{islam2022sharing} require a policy for releasing arithmetic storage held by a stalled requester. Admission-dependent demand can inform this policy. The present validity window governs fast-path use, while product storage remains protected after expiry. Extension to several requesters requires explicit release, completion storage and arbitration, together with a model of their changed ordering.

Window length also determines control structure. Matched registered-mode routes use 6,003 LUTs for unbounded retention and 6,033 at $K=2$, which adds countdown and hold logic. A finite bound therefore carries a control cost. Early admission with one-edge validity provides the measured PaceRV starting point; extending validity recovers residual workload benefit. When finite validity is required, select the smallest window meeting the objective and evaluate its routed cost. Frequency, application cycles and resources jointly determine the implementation choice.

\nobalance
\section{Conclusion}
We studied how memory admission changes the useful retention of prepared arithmetic. Earlier admission in PaceRV removes most delayed opportunities, yet a fractional target can select a longer window. Prospective validation turns this distinction into a design choice: establish admission timing, decide whether residual workload benefit warrants extended validity, then size it against an explicit capture target or missed-opportunity budget. Comparing application cycles and routed cost completes the choice of retention policy.

\bibliographystyle{IEEEtran}
\balance
\def\BIBentryALTinterwordspacing{\begingroup\raggedright}
\def\BIBentrySTDinterwordspacing{\par\endgroup}
\bibliography{references}
\end{document}